\documentclass[aps,prb,twocolumn,showpacs]{revtex4}
\usepackage{amssymb,amsbsy,graphicx,times,subfigure}
\usepackage[latin1]{inputenc}
\newcommand{\be}{\begin{equation}}
\newcommand{\ee}{\end{equation}}
\newcommand{\bea}{\begin{eqnarray}}
\newcommand{\eea}{\end{eqnarray}}

\begin{document}
\title{ Pairing state in the rutheno-cuprate superconductor
   RuSr$\bf_2$GdCu$\bf_2$O$\bf_8$: \\A point contact Andreev reflection
    spectroscopy study}

\author{S. Piano, F. Bobba, F. Giubileo and A. M. Cucolo, \\M. Gombos, A. Vecchione}

\affiliation{Physics Department and INFM-CNR SUPERMAT Laboratory,
University of Salerno, Via S. Allende, 84081 Baronissi (SA),
Italy}

\date{\today}
\begin{abstract}
 The results of Point Contact Andreev Reflection
 Spectroscopy on polycrystalline RuSr$_2$GdCu$_2$O$_8$ pellets are presented.
The wide variety of the measured spectra are all explained in terms
of a modified BTK model considering a \emph{d-wave} symmetry of the
superconducting order parameter. Remarkably low values of the energy
gap $\Delta=(2.8\pm 0.2)meV$ and of the $2\Delta/k_BT_c\simeq 2$
ratio are inferred. From the temperature evolution of the $dI/dV$ vs
$V$ characteristics we extract a sublinear temperature dependence of
the superconducting energy gap. The magnetic field dependence of the
conductance spectra at low temperatures is also reported. From the
$\Delta$ vs $H$ evolution, a critical magnetic field $H_{c_2}\simeq
30 T$ is inferred. To properly explain the curves showing gap-like
features at higher voltages, we consider the formation of a
Josephson junction in series with the Point Contact junction, as a
consequence of the granularity of the sample.
\end{abstract}


\maketitle

\section{Introduction}\label{intro}

Point Contact Spectroscopy \cite{PCS} is a versatile technique
widely used to study the basic properties of superconductors, such
as the density of states at the Fermi level and the superconducting
energy gap. The technique consists in establishing a contact between
a tip of a normal metal (N) and a superconducting sample (S), thus
forming a small contact area that is a ``Point Contact'' junction.
By varying the distance and/or the pressure between tip and sample
it is possible to obtain different tunnel barriers, that is
different conductance regimes. Indeed, quasiparticle tunnel
spectroscopy is obtained for high barriers, while Point Contact
Andreev Reflection (PCAR) spectroscopy is achieved in case of low
barriers. Often in the experiments, intermediate regimes are
realized, in which through the N/S contact both quasi-particle
tunneling and Andreev reflection processes occur.

Andreev Reflections \cite{Andreev,Deutscher} take place at the N/S
interface when an electron, propagating in the normal metal with an
energy lower than the superconducting energy gap, enters in the
superconductor forming an electron pair (Cooper pair) while a hole,
with opposite momentum with respect to the incident electron, is
reflected in the normal metal. A single reflection corresponds to a
net charge transfer of $2e$, where $e$ is the electron charge, from
the normal metal to the superconductor. In the limit of low barriers
at low temperatures, all the incident electrons at the N/S interface
with energy $eV < \Delta$ are Andreev reflected and the conductance
doubles the normal states value.

In this paper we report on PCAR studies carried out in the hybrid
rutheno-cuprate RuSr$_2$GdCu$_2$O$_8$ (Ru-1212) system
\cite{Bauernfeind}. This compound has recently drawn great attention
among theorists and experimentalists in the field of solid state
physics due to the coexistence at low temperatures of
superconducting and magnetic ordering \cite{Review}. The Ru-1212
structure is similar to that of YBa$_2$Cu$_3$O$_7$ with magnetic
(\emph{2D}) RuO$_2$ planes substituting the (\emph{1D}) Cu-O chains.
The superconducting critical temperature in this compound strongly
depends on the preparation conditions with some reports showing
transition onset as high as $50K$ \cite{Bernhard}. The Ru-1212 also
shows a magnetic phase below $135K$. It has been reported that the
magnetic order of the Ru moments is predominantly antiferromagnetic
along the \textsl{c} axis \cite{Lynn}, while a ferromagnetic
component has been observed in the RuO$_2$ planes, that act as
charge reservoir \cite{Bernhard2}. At the moment, due to complexity
of this compound, an exhaustive description of the interaction
between the magnetic and superconducting layers is still missing as
well as an unambiguous evaluation of the symmetry of the energy gap.

 The paper is organized as follows: in Sec.~\ref{model}
we briefly review the results of the BTK model \cite{BTK} for a
conventional \emph{s-wave} superconductor and the recent
extensions \cite{Tanakaart} for anisotropic \emph{s-wave} and
\emph{d-wave} symmetry of the order parameter. In
Sec.~\ref{sample} we describe the point contact experiments in
polycrystalline Ru-1212 pellets showing a variety of conductance
curves obtained at $T=4.2K$. Satisfactory theoretical fittings are
achieved by using a modified BTK model for \emph{d-wave} symmetry
of the order parameter. In Sec.~\ref{jserie} we show that, due to
the granularity of the samples, in some cases, the formation of a
Josephson junction in series with the N/S contact occurs. Indeed,
conductance curves showing gap-like features at higher voltages
and dips in the spectra are well explained by this assumption. In
Sec.~\ref{temperature} we report the temperature evolution of the
conductance curves of a very stable junction. All the spectra are
well reproduced by the \emph{d-wave} modified BTK model, and we
infer from the experiments the temperature dependence of the
superconducting energy gap $\Delta$. In Sec.~\ref{mfd} we analyze
the magnetic field behavior of the measured conductance spectra,
providing an estimation of the upper critical field of the Ru-1212
compound. Finally, in Sec.~\ref{conclu} we summarize our results
and draw some conclusions.

\section{The BTK model and its extension}\label{model}

In this section, for sake of clearness, we review the main results
of the original Blonder-Tinkham-Klapwijk (BTK) theoretical model
\cite{BTK}, as developed for electronic transport through a
Point-Contact junction between a normal metal and a conventional BCS
superconductor. We also summarize the Kashiwaya-Tanaka
\cite{Tanakaart} extension for asymmetric \emph{s-wave} and
\emph{d-wave} superconductors. Indeed, a close comparison of the
calculated conductance spectra is useful for a better understanding
of the peculiar transport processes that occur at an N/S interface
depending on the symmetry of the superconducting order parameter.

\begin{figure}
\includegraphics[width=8.5cm]{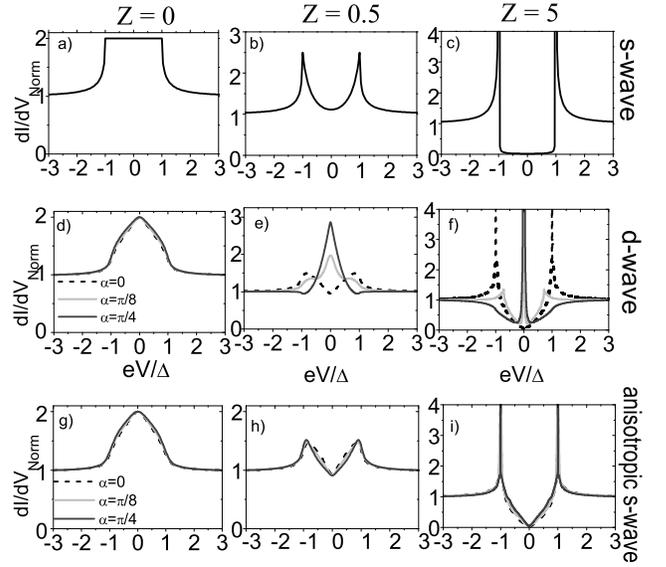}\\
  \caption{Conductance characteristics, at low temperatures, for different
  barriers
  $Z$ as obtained by the BTK model for a Point Contact junction between a
normal metal and a \emph{s-wave} (a,b,c), a \emph{d-wave}(d,e,f) and
an anisotropic \emph{s-wave} superconductor (g,h,i).}\label{all}
\end{figure}

Following the original paper, we write the expression of the
differential conductance characteristics for a N/S contact that,
according to the BTK model\cite{BTK}, is given by:
\begin{eqnarray}\label{BTK}
  \hspace*{-1cm}&&G_{NS}(eV) = \frac{dI(eV)}{dV}= \ \nonumber \\
 \hspace*{-1cm}&&G_{NN}\int_{-\infty}^{+\infty}{dE [1 +
A(E)-B(E)]\left[-\frac{df(E+eV)}{d(eV)}\right]}
\end{eqnarray}
where $eV$ is the applied potential, $G_{NN}=4/(4+Z^2)$ is the
normal conductance expressed in term of $Z$, a dimensionless
parameter modeling the barrier strength,  \emph{f(E)} is the Fermi
function and \emph{A(E)} and \emph{B(E)} are, respectively, the
Andreev reflection and normal reflection probabilities for an
electron approaching the N/S interface. Eq.(\ref{BTK}) shows that,
while ordinary reflections reduce the transport current through the
junction, Andreev reflections increase this by transferring two
electrons (Cooper pair) in the superconducting electrode on the
other side of the barrier. The case $Z=0$ corresponds to a
completely transparent barrier so that the transport current is
predominantly due to Andreev reflections and
$G_{NS}(V<\Delta)/G_{NN}(V>>\Delta)=2$ is found (Fig.\ref{all}a). By
increasing \emph{Z}, the Andreev reflections are partially
suppressed and the conductance spectra tend to the case of a N/I/S
tunnel junction showing peaks at $eV= \pm \Delta$ (Figs.
\ref{all}b,c).

Recently, Kashiwaya and Tanaka \cite{Tanakaart} extended the BTK
model by considering different symmetries of the order parameter.
Indeed, for a \emph{d-wave}
 superconductor, the electron-like and
 hole-like quasiparticles, incident at the N/S interface,
 experience different signs of the order parameter, with
 formation of Andreev Bound States at the Fermi
 level along the nodal directions.
 The presence of Andreev Bound States
 modify the transport current and the expression of the differential conductance is
 given by:
\begin{equation}\label{tanaka}
 G_{NS}(V)=\frac{\int_{-\infty}^{+\infty}{dE\int_{-\frac{\pi}
{2}}^{+\frac{\pi} {2}}{d\varphi \
\sigma(E,\varphi)\cos\varphi\left[-\frac{df(E+eV)}{d(eV)}\right]}}}{\int_{-\infty}^{+\infty}{dE
\left[-\frac{df(E+eV)}{d(eV)}\right]}\int_{-\frac{\pi}
{2}}^{+\frac{\pi}2}{d\varphi \ \sigma_N(\varphi)\cos (\varphi)}},
\end{equation}

\hspace{-0.3cm}where
\begin{equation}\label{tanaka2}
\sigma(E,\varphi)=\sigma_N(\varphi)\frac{(1+\sigma_N(\varphi))\Gamma_+^2+(\sigma_N(\varphi)-1)(\Gamma_+\Gamma_-)^2}{(1+(\sigma_N(\varphi)-1)\Gamma_+\Gamma_-)^2},
\end{equation}

\hspace{-0.3cm} is the differential conductance and

\begin{eqnarray}
\sigma_N(\varphi) &=& \frac{1}{1+\tilde{Z}(\varphi)^2},\quad  \tilde{Z}(\varphi)=Z\cos(\varphi), \\
\Gamma_\pm &=& \frac{E-\sqrt{E^2-\Delta_\pm^2}}{\Delta_\pm},\\
\Delta_\pm &=& \Delta\cos[2(\alpha\mp\varphi)]. \label{tanaka3}
\end{eqnarray}

So, at a given energy $E$, the transport current depends both on the
incident angle $\varphi$ of the electrons at the N/S interface as
well as on the orientation angle $\alpha$, that is the angle between
the \emph{a-axis} of the superconducting order parameter and the
\emph{x-axis}.  When applying Eqs. (\ref{tanaka})-(\ref{tanaka3}) to
PCAR experiments, there is no preferential direction of the
quasiparticle injection angle $\varphi$ into the superconductor, so
the transport current results by integration over all directions
inside a semisphere weighted by the scattering probability term in
the current expression. Moreover, because our experiments deal with
policrystalline samples, the angle $\alpha$ is a pure average
fitting parameter, which depends on the experimental configuration.

In case of \emph{d-wave} symmetry, for $Z\rightarrow 0$, the
conductance curves at low temperatures show a triangular structure
centered at $eV=0$, quite insensitive to variations of $\alpha$ with
maximum amplitude $G_{NS}(V=0)/G_{NN}(V>>\Delta)=2$ (Fig.
\ref{all}d). However, for higher barriers, the conductance
characteristics show dramatic changes as function of $\alpha$. In
particular as soon as $\alpha \neq 0$, the presence of Andreev Bound
States at the Fermi
 level produces strong effects more evident along the nodal direction
($\alpha = \pi/4$) for which $G_{NS}(V=0)/G_{NN}(V>>\Delta)>2$ is
found (Figs. \ref{all}e,f).

For comparison, we report the conductance behavior for anisotropic
\emph{s-wave} superconductor, in which only the amplitude of the
order parameter varies in the $k$-space, while the phase remains
constant and Eq. (\ref{tanaka3}) reduces to:
\begin{equation}\label{tanaka3bis}
\Delta_{+}=\Delta_{-}=\Delta \cos [2(\alpha -\varphi)].
\end{equation}
 Again, in the limit $Z \rightarrow 0$, an
increase of the conductance for $E< \Delta$ with a triangular
profile is found with maximum amplitude
$G_{NS}(V=0)/G_{NN}(V>>\Delta)=2$ at zero bias (Fig.\ref{all}g). On
the other hand, for higher $Z$, we obtain tunneling conductance
spectra that show the characteristic ``V''-shaped profile in
comparison to the classical ``U''-shaped structure found for an
isotropic \emph{s-wave} order parameter (Figs.\ref{all}h,i). We
notice that in this case all the curves are quite insensitive to
variation of the $\alpha$ parameter and a zero bias peak is obtained
only for low barriers.

\section{PCAR SPECTROSCOPY ON RUSR$_2$GDCU$_2$O$_8$: EXPERIMENTS AND THEORETICAL FITTINGS}\label{sample}

 The Ru-1212 samples used for this study were directionally
solidified pellets, grown by means of the Top-Seeded Melt-Textured
method starting from Ru-1212 and Ru-1210 (RuSr$_2$GdO$_6$) powder
mixtures with a ratio Ru-1212/Ru-1210 = 0.2. The details of the
preparation procedure are reported elsewhere. \cite{Attanasio} In
the X-Ray diffraction patterns, a single Ru-1212 phase was found.
In the resistivity measurements versus temperature, the onset of
the superconducting transition was observed at T$_c^{on}\simeq43K$
with $T_c({\rho=0})\simeq24K$ and $\Delta T_c=12K$ ($\Delta T_c$
is defined as the difference between the temperatures measured at
90\% and 10\% of the normal state resistance). We notice that a
broadening of the superconducting transition is often observed in
polycrystalline samples and it is usually related to the formation
of intergrain weak Josephson junctions\cite{Fabrizio, Review,
Cimberle}. We address this point in the next section.

\begin{figure}[tbh]
\includegraphics[width=8.5cm]{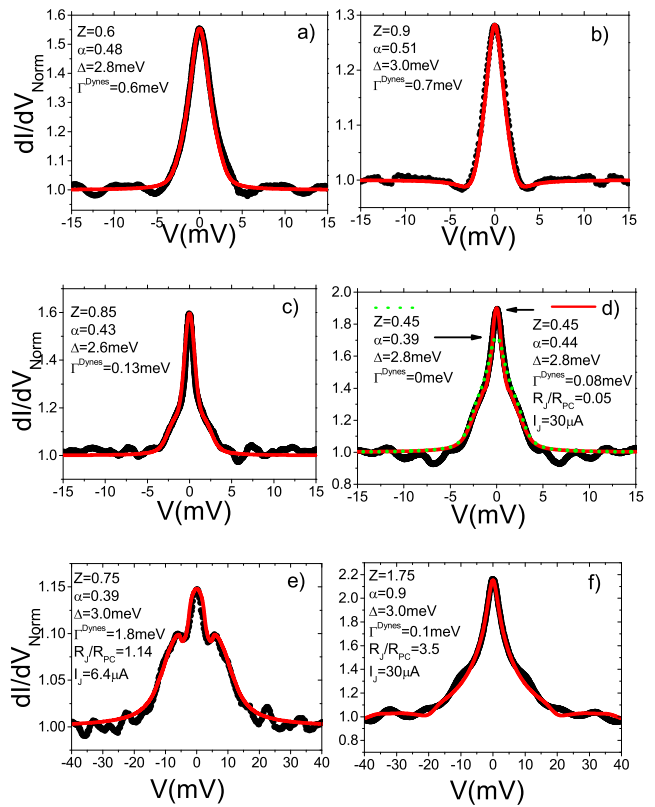}\\
  \caption{(Color online) The $dI/dV$ vs $V$ characteristics measured in different Ru-1212/Pt-Ir
PC junctions at 4.2K. The experimental data (dots) are shown
together with the best theoretical fittings (solid lines) obtained
by a modified BTK model for a \emph{d-wave} symmetry of the
superconducting order parameter.}\label{fit}
\end{figure}

To realize our experiments we used a Pt-Ir tip, chemically etched in
a $40\%$ solution of HCl, while Ru-1212 samples were cleaned in an
ultrasound bath in ethyl alcohol. Sample and tip were introduced in
the PCAR probe, in which three micrometric screws are allocated,
each driven by its own crank. Two screws allow to vary the distance
between tip and sample, with a precision of $1\mu m$ and $0.1 \mu
m$, respectively. The third screw is devoted to change the
inclination of the sample holder varying the contact area on the
sample surface. The Point Contact junctions were formed by pushing
the Pt-Ir tip on the Ru-1212 pellet surface with the probe
thermalized in the liquid He$^4$ bath. The Current-Voltage
 ($I$ vs $V$) characteristics were measured by using a conventional four-probe method
 and a lock-in technique with an amplitude of the ac current less than $1 \mu A$
 was used to measure the differential conductance
 ($dI/dV$ vs $V$) spectra as function of the applied voltage.

 In Fig. \ref{fit}, we show a variety of normalized conductance spectra
 obtained at $T=4.2K$ by establishing different contacts on different areas of the same Ru-1212 pellet.
 The junction resistances
 varied between $10 \Omega$ and $100
\Omega$. By using the Sharvin relation, \cite{Sharvin} it has been
possible to achive an estimation of the size of the contact area.
Indeed $R=\rho l/4a^2$, where $\rho=0.4m \Omega \ cm$ is the low
temperatures resistivity and  $l\approx 1000$ {\AA}, as estimated in
Ref.\cite{Attanasio}. In our case, we have found that the typical
contact size varied between $300${\AA} and $1000${\AA}.

We observe that all the reported spectra are characterized by a Zero
Bias Conductance Peak (ZBCP) with a triangular structure, the main
features appearing for each contact with different shapes,
amplitudes and energy widths. Quite often, oscillations are observed
on the conductance background, as shown in Figs. \ref{fit} c--f. We
observe that the ZBCP appears as a simple structure in Fig.
\ref{fit}b, while in the remaining spectra it results to be
structured with variations of slope or secondary maxima, as in Fig.
\ref{fit}e. The maximum conductance ratio
$G_{NS}(V=0)/G_{NN}(V>>\Delta)$ is less than $2$ for all the curves,
however $G_{NS}(V=0)/G_{NN}(V>>\Delta)\simeq 2.2$ for the data in
Fig. \ref{fit}f. In addition to this, the energy width of the main
zero bias triangular structure is lower than $10mV$ in Figs.
\ref{fit}a--d while it results wider, around $40mV$, in Figs.
\ref{fit}e,f. At a first qualitative analysis, these data appear
quite puzzling and could be interpreted in term of local, large
variations of the superconducting energy gap. In the following, we
will show that the theoretical fittings of all the spectra give
clear indication of a \emph{d-wave} symmetry of the superconducting
order parameter, with consistent values of the inferred amplitude of
the energy gap.

First of all, let us quantitatively analyze the curves of Figs.
\ref{fit}a--d.  We were not able to reproduce the conductance
spectra reported in Fig. \ref{fit} a,b by using either the
conventional \emph{s-wave} model or the anisotropic one, even by
considering small Z values, indicative of low barriers. On the
other hand, as can be observed in Fig. 1, the \emph{s-wave}
fittings cannot model the structured conductances reported in
Figs. \ref{fit} c,d. The solid lines in the figures are the
theoretical fittings obtained by considering a \emph{d-wave}
symmetry of the order parameter in the modified BTK model, Eqs.
\ref{tanaka}-\ref{tanaka3}. A satisfactory agreement is obtained
by using as fitting parameters the superconducting energy gap
$\Delta$, the barrier strength $Z$, the angle $\alpha$ and a
phenomenological factor $\Gamma^{Dynes}$ \cite{Dynes} to take into
account pair breaking effects and finite quasiparticle lifetime
\cite{gamma}. We notice that in the considered spectra, both
quasiparticle tunneling and Andreev reflection processes take
place, since intermediate $Z$ values have to be used to simulate
the barrier strength ($0.45\leq Z \leq 0.9$). Moreover, the angle
$\alpha$ varies between $0.39$ and $0.51$, indicating that the
average transport current mainly flows along an intermediate
direction between the nodal one ($\alpha = \pi/4$) and that of the
maximum amplitude of the energy gap ($\alpha = 0$). The modified
{\em d-wave} BTK model allows to satisfactorily reproduce the
variations of slope around $\pm 1mV$ of the structured ZBCP in
Figs. \ref{fit}a,c,d with a light discrepancy in modeling the full
height of the peak in Fig. \ref{fit}d.  We show in the next
section that a more satisfactory fitting for this contact can be
obtained by taking into account an additional in series intergrain
junction.

We remark that the values of the superconducting energy gap,
inferred from the theoretical fittings, are all consistent and
enable us to estimate an average value of the amplitude of the
order parameter $\Delta=(2.8\pm 0.2)\ meV$. This value is
surprisingly low in comparison with the amplitude of the energy
gap in other cuprate superconductors, however the possibility that
the presence of the RuO$_2$ magnetic planes can play an important
role in the complex Ru-1212 system has to be taken into account.
We notice that the ratio between the smearing factor
$\Gamma^{Dynes}$ and the superconducting energy gap results always
less than $20\%$ and it vanishes for the fitting shown in Fig.
\ref{fit}d. We consider this fact as an indication of the good
quality of our point-contact junctions.

\section{Role of the intergrain coupling}\label{jserie}

\begin{figure}[b!]
\includegraphics[width=5cm]{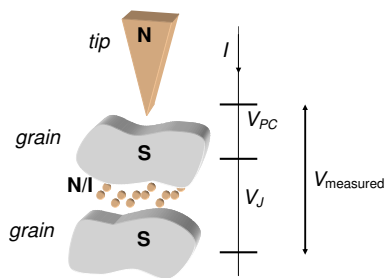}\\
  \caption{(Color online) Intergrain coupling effect in polycrystalline samples. The
  measured voltage $V_{measured}$ is the sum of two terms: $V_{PC}$, the voltage drops between tip and
  sample, the N/S PC junction, and $V_J$, the voltage drops between two superconducting
  grains, forming the S/I/S Josephson junction.}\label{intergrain}
\end{figure}
  To complete our discussion about the spectra measured at low
  temperatures, we now address the analysis of the conductance curves reported in Figs.
  \ref{fit}e,f, with a wider ZBCP. In this respect, we observe that, due to the granularity of the
compound, in same cases, an intergrain Josephson junction can be
formed in series with the Point Contact one, as schematically drawn
in Fig. \ref{intergrain}. This topic has been recently addressed in
PCAR studies on MgCNi$_{3}$ \cite{cinesi} and MgB$_2$ \cite{Filgiu}.

To provide a quantitative evaluation of the conductance spectra,
we consider a real configuration in which the Pt-Ir tip realizes a
PC junction on a single Ru-1212 grain, which, in turn, is weakly
coupled to another grain, so forming a Josephson junction.    In
this case the measured voltage corresponds to the sum of two
terms:
 \begin{equation}\label{lee}
V_{\rm measured}(I)=V_{PC}(I)+V_J(I)\;,
  \end{equation}
where $V_{PC}$ and $V_J$ are the voltage drops at the N/S Point
Contact junction and at the S/I/S intergrain Josephson junction,
respectively. This last contribution can be calculated by the Lee
formula \cite{Lee} which, in
  the limit of small capacitance and at low temperatures, reduces to the simplified
  expression \cite{Duzer}:
\begin{equation} \label{samjose}
V_{J} =
\left\{%
\begin{array}{cl}
    0 & \hbox{\quad for $I<I_{J}$ ;} \\
    R_J I_{J}\sqrt{[(I/I_{J})^2-1]} & \hbox{\quad for $I \ge I_{J}$ .} \\
\end{array}%
\right.
\end{equation}
At the same time, for the Point Contact contribution, we use again
the extended BTK model for a \emph{d-wave} superconductor. The
$I(V)$ characteristic is then calculated by inverting Eq.
(\ref{lee}) and the conductance spectrum is given by:
\begin{equation}
\sigma(V)=\frac{dI}{dV}=\left(\frac{dV_{PC}}{dI}+\frac{dV_J}{dI}\right)^{-1}.
  \end{equation}

By applying this simple model we have satisfactory fitted the
experimental data reported in Figs. \ref{fit}e,f. Remarkably, for
both spectra, the best fittings have been obtained by using
$\Delta=3.0 \ meV$, consistently with the average value extracted
from the other curves in Figs. \ref{fit}a--d. We observe that, in
this model, two more parameters are needed, namely the resistance
$R_J$ and the critical current $I_J$ of the Josephson junction.
However, the choice of these two parameters is not completely
arbitrary, since the condition $R_J+R_{PC}=R_{NN}$ has to be
fulfilled, where $R_{NN}$ is the measured normal resistance and
the product $R_JI_J$ necessarily results lower than $\Delta$
\cite{Barone}.

In some cases, it has been pointed out that dips in the
conductance spectra can be related to the presence of intergrain
junctions \cite{cinesi, Filgiu}, and for sake of completeness, we
have applied our model also to the spectra of Figs. \ref{fit}
a--d. We notice that for different junctions, the effect of the
intergrain coupling results more or less evident, depending on
ratio $R_J/R_{PC}$. For the conductances shown in Figs.
\ref{fit}a--c this effect turns out to be negligible, however,
some improvement of the theoretical fitting is obtained in the
case of Fig. \ref{fit}d (see dashed line). Remarkably, by this
last fitting we have found the same value of the superconducting
energy gap previously inferred, $\Delta=2.8 meV$, with a
$\Gamma^{Dynes} / \Delta$ ratio less than $3 \%$ and
$R_J/R_{PC}=0.05$.
\smallskip

\section{Temperature dependence of the conductance spectra}\label{temperature}
\begin{figure}[t!]
\includegraphics[width=8.5cm]{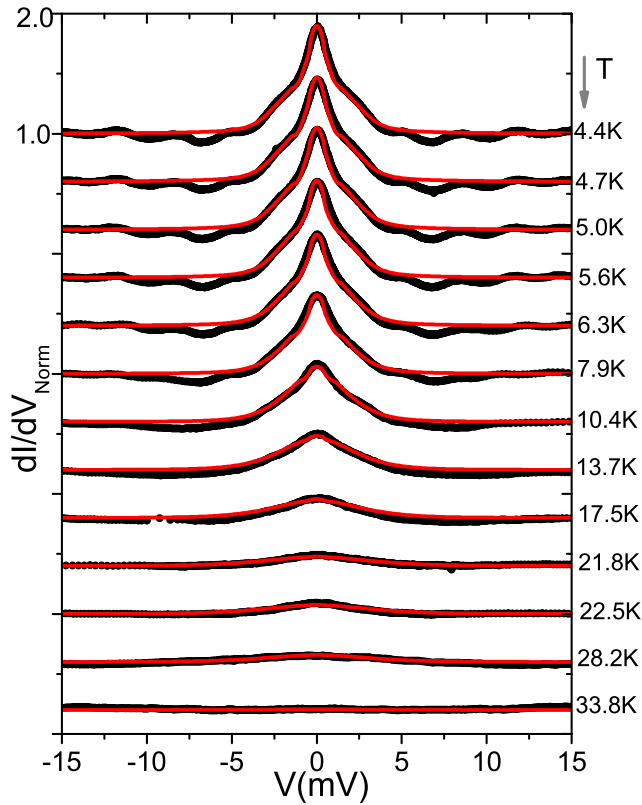}\\
  \caption{(Color online) Temperature evolution of the conductance spectrum of Fig. \ref{fit}c from $T=4.2K$ to up the critical
  temperature (dots). The solid lines are the theoretical fittings obtained
  by a
   modified \emph{d-wave} BTK model with the energy gap as only free parameter.}\label{temperatura}
\end{figure}
\begin{figure}[t!]
\includegraphics[width=8cm]{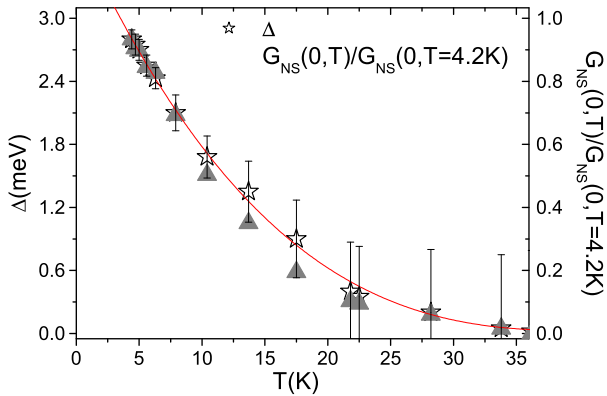}\\
  \caption{(Color online)Temperature dependence of the superconducting energy gap as inferred from the theoretical fittings
  shown
  in Fig. \ref{temperatura}. The solid line is a guide for the eyes. The
 right hand scale refers to the temperature evolution of the
 measured height of the ZBCP normalized to the 4.2K value.
}\label{delta}
\end{figure}

To achieve information on the temperature dependence of the
superconducting energy gap in the Ru-1212 system, in this section we
analyze the temperature behavior of the conductance spectrum shown
in Fig. \ref{fit}d. Indeed, this PCAR junction resulted to be very
stable for temperature variations.

In Fig. \ref{temperatura} we show the conductance characteristics
measured in the temperature range $4.2K \le T < 35K$. We firstly
notice that the ZBCP decreases for increasing temperature and
disappears at about $T\simeq 30K$, that we estimate as the local
critical temperature $T_c^l$ of the superconducting Ru-1212 grain
in contact with the Pt-Ir tip, consistently with the resistivity
measurements\cite{Attanasio}. This fact provides further evidence
that the ZBCP is a consequence of the superconducting nature of
Ru-1212 and is not due to spurious effects like inelastic
tunneling via localized magnetic moments in the barrier region
\cite{Cucolo}. The experimental data for each temperature are then
compared to the theoretical fittings calculated by using the
\emph{d-wave} modified BTK model with a small contribution of
Josephson junction in series. For all the curves, we fixed the
strength of the barrier and the angle $\alpha$ to the values
obtained at the lowest temperature.

The resulting temperature dependence of the superconducting energy
gap $\Delta(T)$ is reported in Fig. \ref{delta}, where vertical
bars indicate the errors in the gap amplitude evaluation, that
increase when approaching the critical temperature. Contrarily to
what expected for BCS superconductors, we observe that the energy
gap, at low temperatures, decreases rapidly for increasing
temperatures and goes to zero at $T_c^l$ in a sublinear way. We
notice that the same temperature evolution for the superconducting
energy gap is found trough the \emph{d-wave} BTK model with or
without considering any intergrain junction in series; remarkably,
in this last case, the superconducting energy gap $\Delta$ remains
the only varying parameter.  A similar temperature dependence has
been reported by G. A. Ummarino \emph{et al.}\cite{Gonnelli},
however these authors give a larger estimation of the maximum gap
amplitude.

 From the average value of the superconducting energy gap
$\Delta=2.8\ meV$ and from the measured local critical temperature
$T_c^l\simeq 30K$, we obtain a ratio $2\Delta/(k_B T_c^l)\simeq2$
much smaller than the predicted BCS value and also smaller than the
values found for high-$T_c$ cuprate superconductors\cite{YBCO}.
Again we speculate that the simultaneous presence of superconducting
and magnetic order is an important key for understanding the
behavior of the Ru-1212 system. Coexistence of superconductivity and
antiferromagnetism is found among cuprates, however it is common
believe that ferromagnetism and superconductivity are mutually
excluding orders. Recently, it has been found that in conventional
Superconductor/Ferromagnetic (S/F) structures, proximity effect give
rise to an oscillatory behavior of the superconducting $T_c$ as a
function of the thickness of the F layer
\cite{proximity,proximity2}. There are conditions for which a change
of sign of the order parameter occurs, producing the $\pi$-junction
phenomenon\cite{pijunction}. In addition to this, a dramatic
suppression of the amplitude of the order parameter is expected for
high $T_c$ superconductors in close contact with a ferromagnetic
material \cite{YBCO_LCMO} and various examples of anomalous
temperature behavior are found in the literature. Gapless
superconductivity can be achieved, that can induce a sublinear
temperature dependence of the superconducting energy gap. In the
Ru-1212 system, it has been proved that the RuO$_2$ planes are
conducting, however these do not develop superconductivity at any
temperature\cite{Pozek}. By means of different experimental
techniques, it has been inferred that a large fraction of the charge
carriers is not condensed in the superconducting state even at low
temperatures \cite{Pozek}. Both findings are consistent with a
reduced value of the $2\Delta/(k_B T_c)$ ratio in this compound.

In Fig. \ref{delta} we also report (righthand scale) the temperature
evolution of the height of the ZBCP normalized to its value at
$T=4.2K$. It is worth to notice that $G_{NS}(V=0,T)$, as directly
measured from the experiments, and $\Delta(T)$, as inferred from the
theoretical fittings, show a similar scaling with temperature. This
correspondence is easily verified for $Z=0$ in case of a
\emph{s-wave} superconductor, however it is a quite new result since
it has been found for intermediate barriers and unconventional
symmetry of the superconducting order parameter.

\section{MAGNETIC FIELD DEPENDENCE OF THE CONDUCTANCE SPECTRA}\label{mfd}

As we already observed, one of the most interesting features of the
Ru-1212 is the coexistence of the superconducting phase and magnetic
order. Indeed, from Nuclear Magnetic Resonance (NMR) \cite{NMR,NMR2}
and magnetization \cite{magnetization} measurements, it has been
found that in this compound ruthenium occurs in a mixed valence
state $Ru^{4+}$, $Ru^{5+}$ with some higher $Ru^{5+}$ concentration.
\begin{figure}[b!]
 \includegraphics[width=8.5cm]{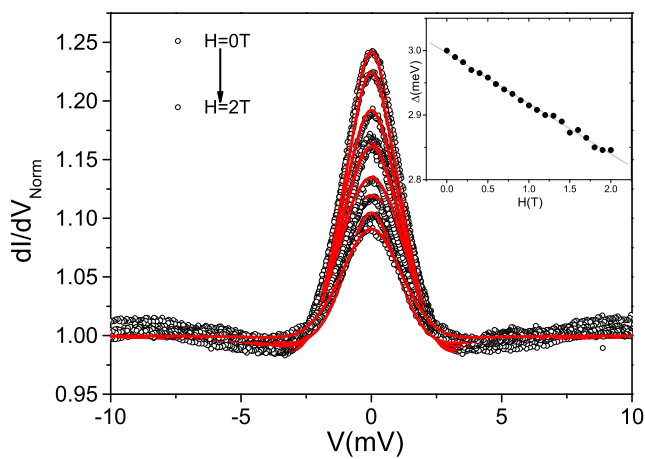}\\
  \caption{(Color online) Magnetic field dependence of the normalized $dI/dV$ vs $V$ characteristics at $T=4.2$ K from $0$ T to $2$ T
  (dots) for the spectra of Fig. \ref{fit}b. The full lines are the theoretical fittings obtained as
  discuss in the text. In the inset the magnetic field dependence of
  the energy gap is reported.}\label{camponew}
\end{figure}
The $RuO_2$ planes, from one side, act as charge reservoir for the
superconducting $CuO_2$ planes, on the other hand, as observed in
Muon Spin Rotation ($\mu$SR) experiments \cite{Bernhard2}, they show
quite homogeneous ferromagnetic order below $T_c$. A weak
interaction between the two order parameters, ferromagnetism in the
$RuO_2$ planes and superconductivity in the $CuO_2$ planes, has been
suggested and recently several experiments  appear to confirm this
hypothesis \cite{Pozek}. Despite of the huge experimental and
theoretical efforts focused on the study of the interplay between
superconductivity and magnetism, to the best of our knowledge no
spectroscopic studies in magnetic field of the superconducting order
parameter in Ru-1212 have been reported in literature so far.

In Fig. \ref{camponew} we show the PCAR spectra measured by applying
an external magnetic field, parallel to the tip, with intensity $H$
varying from $0T$ to $2 T$. The $dI/dV$ vs $V$ curves refer to the
contact reported in Fig. \ref{fit}b. A reduction of the ZBCP for
increasing magnetic fields is observed, that in first approximation
can be reproduced by a phenomenological approach. Indeed, addressing
the problem of the magnetic field dependence of the conductance
characteristics is non conventional superconductors, is a quite
difficult task and a complete treatment of PCAR spectroscopy in
magnetic field would require the use of an appropriate density of
states in calculating the BTK expression for the reflection and
transmission coefficients at the N/S interface. Due to the lack of
an analytical model, Miyoshi \emph{et al.}\cite{Yuri} presented a
two fluid model to reproduce the effect of normal vortex cores in
PCAR junctions in conventional superconductors, assuming that the
contact area contains multiple randomly distributed individual
junctions (non-Sharvin regime \cite{Sharvin}). These authors propose
a simplified expression for the conductance, written as a sum of
normal and superconducting channels:
$$G_{NS_{tot}}(V)=(1-h)G_{NN}+hG_{NS}(V)$$ where $h=H/H_{c_2}$ and $H_{c_2}$ is the critical field. This approach, however,
cannot be applied to our experiments since we deal with
policrystalline, unconventional superconductor, exhibiting internal
magnetic ordering. In this case, the magnetic induction $B$ is not
simply proportional to the external magnetic field $H$ and as a
consequence the density of vortices is not linearly related to $H$.

An alternative way to perform a theoretical fitting is obtained by
using an additional pair breaking parameter to simulate the effect
due to the magnetic field \cite{Naidyuk,Gonnelli2}. In this case,
the total broadening effect $\Gamma$ is considered as the sum of two
terms: $\Gamma=\Gamma^{Dynes}+\Gamma^{ext}$ where $\Gamma^{Dynes}$
is the intrinsic broadening due to the quasiparticle lifetime, as
used in the previous fittings, while $\Gamma^{ext}$ mimics the pair
breaking effect due to the external applied magnetic field. The
curve at $H=0$T (see Fig. \ref{fit}b) has been fitted by using the
\emph{d-wave} modified BTK model with $\Delta=3.0\ meV$. For
increasing magnetic fields we keep constant, in the numerical
computation, the strength of the barrier $Z=0.9$, the orientation
angle $\alpha=0.51$ and the intrinsic $\Gamma^{Dynes}=0.7\ meV$,
while varying only two parameters: the energy gap $\Delta$ and the
magnetic field effect $\Gamma^{ext}$. We observe that the best
theoretical fittings (solid lines in Fig. \ref{camponew})
satisfactorily reproduce for any field both the height and the
amplitude of the measured spectra. In the inset, we report the
magnetic field dependence of the superconducting energy gap (dots)
as extracted from the theoretical fittings. The amplitude of the
energy gap reduces linearly for $H$ up to 2T and by a linear
extrapolation of the data, we find that the energy gap disappears at
about $H^{ext}\simeq 30$ T, consistently with the estimated critical
field reported in Ref.\cite{Attanasio}.

\begin{figure}
\includegraphics[width=8.5cm]{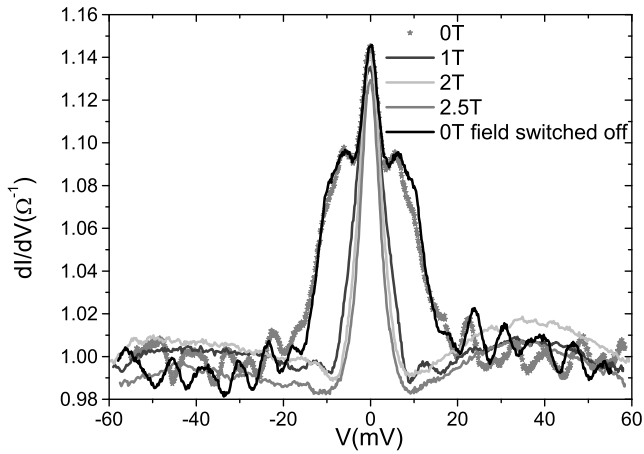}\\
  \caption{Normalized conductance curves for the contact of Fig. \ref{fit}e measured at
$T=4.2K$ in magnetic field up to 2.5 T. When the field is switched
off, the original spectra are recovered.}\label{serie}
\end{figure}

We have also studied the effect of the magnetic field on the
conductance characteristics of the junctions showing wider ZBCP,
that are formed by two junctions in series. In Fig. \ref{serie} we
report the $dI/dV$ vs $V$ curves measured up to $2.5$ T for the
contacts of Fig. \ref{fit}e. In this case, we observe that the
conductance curves dramatically change with the application of the
magnetic field. As discussed in the previous section, the Josephson
current due to the intergrain coupling is immediately suppressed by
the magnetic field, modifying the spectra towards the narrower,
non-structured, triangular shape of the ZBCP. In addition to this,
the oscillatory behavior of the background, due to the intergrain
coupling disappears in magnetic field. Remarkably, for the junctions
of both Figs. \ref{camponew},\ref{serie} the peculiar features of
the spectra together with the normal junction resistance, are
recovered when the magnetic field is switched off, and no hysteresis
is found for increasing/decreasing fields.

\section{Conclusions}\label{conclu}
We have analyzed the PCAR conductance spectra obtained in
superconducting RuSr$_2$GdCu$_2$O$_8$ (Ru-1212) policrystalline
pellets. All the conductance curves at low temperatures show a Zero
Bias Conductance Peak that decreases for increasing temperatures and
disappears at the local critical temperature $T_c^l\simeq 30K$ of
the superconducting grain in contact with the Pt-Ir tip. The
triangular shape of all the measured spectra has been modeled by
using a modified BTK model for a \emph{d-wave} symmetry of the
superconducting order parameter. This finding suggests a closer
similarity of the Ru-1212 system to the high $T_c$ cuprate
superconductors rather than to the magnetic ruthenate Sr$_2$RuO$_4$
compound. However, the remarkably low values of the energy gap
$\Delta=(2.8 \pm 0.2) meV$ and of the ratio $2 \Delta/k_BT_c\simeq2$
indicate major differences between the Ru-1212 and the high $T_c$
cuprates. We speculate that the presence of ferromagnetic order
within the superconducting phase results in an effective reduction
of the energy gap. We have also demonstrated that, when dealing with
granular samples, intergrain coupling effects can play a predominant
role. In some cases, an intergrain Josephson junction in series with
the point contact junction is formed. Taking into account this
feature as well, all conductance spectra have been properly modeled
by considering a {\em d-wave} symmetry of the order parameter, with
consistent values of the amplitude of the energy gap.

By fixing all the fitting parameters to their values at the lowest
measured temperature, and by varying $\Delta$, the temperature
dependence of the energy gap has been extracted from the conductance
characteristics of a very stable junction. We have found that the
energy gap exhibits a sub-linear dependence in temperature. The
magnetic field behavior of the spectra has been also studied,
showing a linear reduction of the energy gap for fields up to $2$ T,
from which a critical field $Hc_2\sim 30$ T is inferred. We have
found that both the superconducting features and the normal
background in the conductance spectra do not show any hysteresis in
magnetic field. These observations seem to suggest a
 weak coupling between the superconducting and magnetic order
parameter.

Our analysis may be helpful for a deeper understanding of the
mechanisms enabling high temperature superconductivity, and its
interplay with magnetic order in unconventional superconductors like
rutheno-cuprates.

 \acknowledgments{The authors thank Y. Maeno, Y. Tanaka and G. Deutscher for helpful discussions and F. Vicinanza for the technical support.}

\end{document}